# G315.1+2.7: a new Galactic SNR from the AAO/UKST Hα survey

M. Stupar,[1] Q.A. Parker,[1,2] M.D. Filipović[3]

[1]*Department of Physics, Macquarie University, Sydney, NSW 2109, Australia*
[2]*Anglo-Australian Observatory, P.O. Box 296, Epping, NSW 1710, Australia*
[3]*University of Western Sydney, Locked Bag 1797, Penrith South DC, NSW 1797, Australia*

**ABSTRACT**

New narrow-band Hα imaging and subsequent optical spectra confirm G315.1+2.7, a previously identified candidate supernova remnant (SNR), as a bona-fide Galactic SNR. Present observations are based on independent discovery of filamentary optical emission nebulosity on images of the AAO/UKST Hα survey of the southern Galactic plane (Parker et al. 2005) that were found to coincide with existing multi-frequency radio detections. Separate medium and high dispersion spectra were taken across two locations of this 11 arcmin N-S aligned optical filament. The resulting spectral signatures were found to strongly confirm the SNR identification based on standard emission line ratio discriminators which characterise emission from shock heated gas (Fesen, Blair, Kirshner 1985). The average observed ratios of [SII]/Hα=1.13, [NII]/Hα=1.43 and [SII] 6717/6731=1.46, together with the simultaneous detection of [OII] at 3727Å, [OIII] at 5007Å and [OI] 6300Å, all point to an SNR origin of the observed optical emission.

There is also excellent positional coincidence between the new Hα filament and the northeast radio arc of G315.1+2.7 seen at several frequencies. Careful scrutiny of the low-resolution but high sensitivity SHASSA Hα survey (Gaustad et al. 2001) also revealed a low-level but distinct optical emission arc. This arc precisely correlates with the large, 2.5 degree, north-south angular extent of the proposed new SNR also seen as a fractured structure in the extant radio data. G315.1+2.7 was detected previously at 2400 and 4800 MHz (Duncan et al. 1995, 1997) and at 408 and 1420 MHz (Combi, Romero & Arnal 1998). We also identified associated radio emission at 843 MHz from the now publicly available SUMSS survey. On the basis of optical imaging and spectra and radio observations at five frequencies, we identify G315.1+2.7 as a new Galactic SNR. The large projected angular extent of the new remnant, together with the distance estimate of ~1.7 kpc and diameter of ~80 pc, make G315.1+2.7 one of the largest remnants known.

## 1 INTRODUCTION

Historically, Galactic SNRs and their complex filamentary structures were firstly discovered in optical images (e.g. the Vela SNR and Cygnus Loop). Today however, the large majority of Galactic SNR's have been discovered and confirmed from multi-frequency radio observations where the observed spectral indices betray their non-thermal nature. All 231 Galactic SNRs currently catalogued by Green (2004) have multi-frequency radio confirmation whereas only ~ 20% of known SNRs have detected optical counterparts and only 15 % have associated optical spectra.

The inhomogeneity of interstellar space means that optical emission does not arise directly from the primary shock for every SNR. The small percentage of current optically detected Galactic SNRs are usually first seen in Hα light as the powerful SNR blast wave propagates into a cool and dense ISM leading to shock excited radiation at the interface. Apart from HAα emission, other Balmer lines such as Hβ and Hγ are often detected. Forbidden lines are also seen such as [SII] 6717/6731Å, [NII] 6548/6584Å, [OI] 6300/6364Å, [OIII] 4959/5007Å and [OII] 3727Å. The [SII] lines can be particularly strong and a ratio of [SII] to Hα >0.5 is used as a standard discriminator for shock excited gas and hence indicative of an SNR. Fesen, Blair & Kirshner 1985) provide useful diagnostic line-ratio diagrams which are effective in separating HII regions and planetary nebulae (PN) from the shock emission of SNRs.

Galactic optical SNRs in Hα light usually have distinct morphological characteristics in the form of narrow filaments and other complex, intertwined and elongated arc structures. During systematic visual inspection of the original survey films of the AAO/UKST Hα survey of the southern Galactic plane (Parker 2005), an isolated filamentary nebulosity, extending for ~11 arcmin in a N-S direction, was identified at R.A.=$14^h 33^m 16^s$ δ=-57° 32' 40" (J2000). The filament's position was checked for prior optical detection in SIMBAD (nothing noted) and against available radio survey data. A very good positional coincidence was found with radio emission at several frequencies, and in particular with the latest Sydney University Molonglo Sky Survey (SUMSS) images at 843 MHz (Bock, Large & Sadler 1999).[1]

Duncan et al. (1995, 1997) originally reported a shell SNR candidate with a diameter of 2.5°✕3.2° at the same general location as the new optical filament based on radio emission at 2400 MHz and on the corresponding Parkes-MIT-NRAO (PMN) radio survey data at 4850 MHz (Griffith & Wright 1993). Here it is shown that the new Hα filament is clearly associated with the radio detections and that we have uncovered the optical counterpart of this candidate SNR that Duncan (1995) designated as G315.1+2.7.

The preliminary spectral observations of the Hα filament and the detection of prominent spectral lines which characterize the emission as arising from an SNR are reported. We connect our observations with previous radio continuum observations at several different frequencies that provide an estimated spectral index (Combi, Romero & Arnal 1998) and confirm the non-thermal nature of the radio emission from G315.1+2.7. Taken together, all the evidence now points strongly to the identification of G315.1+2.7 as a new Galactic SNR.

## 2 OPTICAL OBSERVATIONS

### 2.1 Detection of the Hα emission

The high (≤5 Rayleigh) sensitivity, wide area coverage (4000~sq. degrees) and arcsecond resolution of the AAO/UKST Hα[2] survey of the southern Galactic plane (Parker et al. 2005) provides a new opportunity to search for faint, low surface brightness emission nebulosities. These include HII regions (Cohen et al. 2002), Planetary nebulae (Parker et al. 2003, 2006), and new Galactic SNRs (e.g. Parker, Frew & Stupar 2004, Stupar 2006). For SNRs in particular, searches for large-scale, correlated, diffuse emission and/or distinct filamentary structures were systematically performed by MS for his thesis project.

Visual inspection of all 233 fields of the Hα survey was first undertaken by MS, QAP and David Frew (looking for extended planetary nebulae and Galactic SNRs) on the 16✕16 blocked-down FITS images of each field available on line from the general Hα survey web site[3]. Each blocked-down survey field has an accurate World-Coordinate System (WCS) built in to the FITS header enabling accurate positions of any newly identified emission cloud or filament to be recorded for later download of the full (0.67 arcsecond/pixel) resolution SuperCOSMOS Hα Survey (SHS) data. Following this examination and subsequent rejection of known HII regions and large PN independently identified and verified by Frew & Parker (2005), about 50 new large-scale emission nebulosities of various sizes were found. However, it is important to note that the 16✕16 blocked fields, although excellent for improving the detection of coherent, large-scale, low-surface brightness nebulosities, are less sensitive to the detection of isolated or fragmented filamentary structures due to the highly suppressed resolution. Hence, any isolated fine-structure Hα filaments (from tens of arcseconds to several arminutes in size) may be lost (see Parker, Frew & Stupar 2004 for such example). Ideally, the searches of the blocked-down data should be supplemented by complimentary scrutiny of the full resolution Hα survey imagery. However, as the survey covers 4000 square degrees, downloading the full resolution digital data for the entire survey from the SHS web site is not practical, notwithstanding the limitation of a maximum download image size of 30✕30 arcmin at any one time. Hence a systematic visual inspection of the original survey films was undertaken at the Plate Library of the Royal Observatory Edinburgh in August 2004 by MS looking specifically for large-scale filamentary structures. This led to the tures. This led to the discovery of an additional 60 such filaments not seen in the 16× blocked fields. All such newly identified Hα filaments could be possible tracers of new optically detected Galactic SNRs. Overall results will be presented in Stupar, Parker & Filipović (in preparation) and in other papers in the series. Here we present specific details concerning one of these newly discovered optical filaments associated with the SNR candidate G315.1+2.7.

---

[1] http://www.astrop.physics.usyd.edu.au/SUMSS/

[2] also covering [NII] emission at 6548 and 6584Å.
[3] http://www-wfau.roe.ac.uk/sss/halpha/

## 2.2 Optical identification of G315.1+2.7

Although G315.1+2.7 can be seen as an elongated nebular "blob" in the blocked field data, the filament was discovered from visual scrutiny of the original Hα survey film in Edinburgh Observatory Plate Library. The true filamentary nature of the structure is only evident in the high resolution image data where it is seen to extend in a N-S direction for ~11 arcmin with a typical width of ~0.8 arcmin. Figure 1 shows the structure and full extent of this new filament taken from the SHS data.

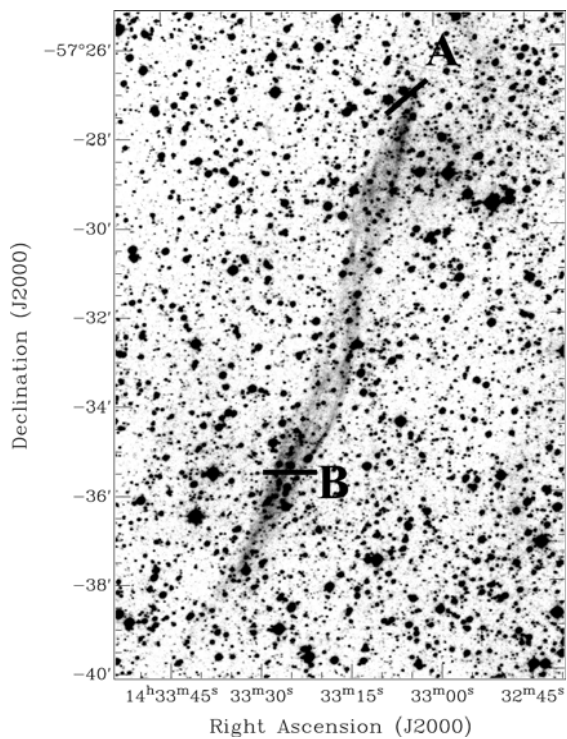

Fig. 1 The AAO/UKST Hα image of the G315.1+2.7 optical filament. The positions and orientations of the 2.5 arcsec Double Beam Spectrograph (DBS) slit are given (A and B). For slit position coordinates see Table 1.

As a supplementary discovery medium to the high resolution SHS, the Southern Hα Sky Survey Atlas (SHASSA) of Gaustad et al. (2001) provides an additional source of lower-resolution but higher sensitivity Hα imaging. Each SHASSA field covers 13°×13° at 48 arcsec resolution. Careful examination of the SHASSA data in the vicinity of G315.1+2.7 reveals the full optical extent of the SNR. There is a very low-level Hα emission structure that can just be seen as an extensive arc structure running from north to south over 2.5° (Fig. 2). The additional arrows on this figure highlight the position of the arc. The short filament identified from the SHS is indicated with a nearly vertical black curly line to the east. To the west in this image is part of a huge interstellar Hα emission structure, which extends more than 10 degrees from the south-west part of G315.1+2.7.

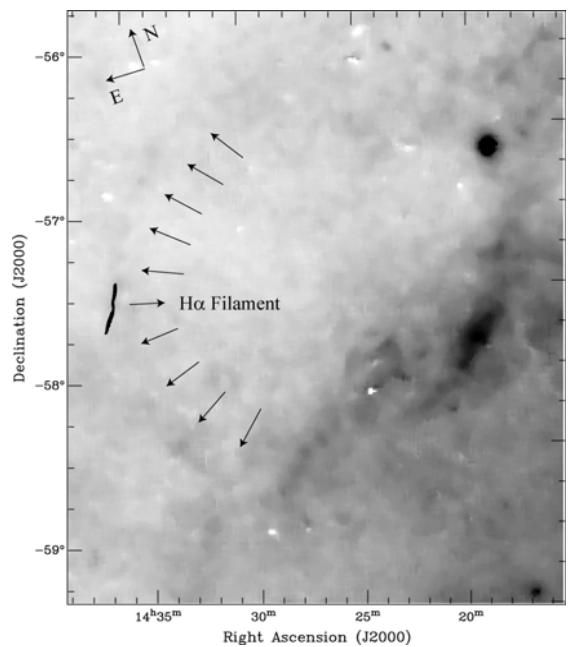

Fig. 6.2 SHASSA image of the area around G315.1+2.7. Arrows show the position of the weakly Hα emitting north-south shock front, which is in excellent positional coincidence with the radio arc(s). The brighter structures to the South and West are unrelated Galactic emission.

## 2.3 Medium and High resolution optical spectra

The optical spectral observations presented here (see Table 1) were made with the double beam spectrograph (DBS) of the 2.3m telescope on Mount Stromlo and Siding Spring Observatory (MSSSO) on June 3 and June 6, 2005 for slit positions A and B respectively. The 2.5 arcsecond wide spectrograph slit was positioned and oriented across the filament at the two locations approximately 9 arcminutes apart and labelled A and B on Fig. 1. For both nights the same exposure time of 1200 sec was used. The blue arm of the DBS was equipped with a 600 lines/mm medium dispersion grating angled to cover the wavelengths from 3500 to 5400Å and an instrumental resolution of 2.2Å (1.1Å/pixel). The red arm of the DBS was equipped with a high resolution 1200 lines/mm grating covering about 800Å between 6100 to 6900Å. The instrumental resolution was 1.1Å and the beam is split into the red and blue arms by a dichroic.

Table 1. Summary of spectroscopic observations of the Hα filament of G315.1+2.7.

| Date | DBS[a] arm | Slit R.A. (h m s) | Slit δ (° ′ ″) | Wavelength Range (Å) | Slit position |
|---|---|---|---|---|---|
| 03/06/2005 | blue | 14 33 25 | -57 35 30 | 3500-5400Å | A |
| 03/06/2005 | red  | 14 33 25 | -57 35 30 | 6100-6900Å | A |
| 06/06/2005 | blue | 14 33 05 | -57 27 40 | 3500-5400Å | B |
| 06/06/2005 | red  | 14 33 05 | -57 27 40 | 6100-6900Å | B |

[a] Blue grating=600 lines/mm; Red grating=1200 lines/mm
All exposure times were 1200 seconds

Standard slit-spectra IRAF reduction routines were used. For absolute flux calibration two standard photometric stars were observed: LTT 3864 on June 03, 2005, and LTT 7379 on June 06. Furthermore, additional routines from the Starlink[4] software package FIGARO were used as well as supplementary IRAF scripts developed by Brent Miszalski, Macquarie University, Sydney, (private communication) that facilitated faster extraction of 1-D spectra.

Examination of the flux-calibrated spectra, especially the red part, confirms classification of this object as a SNR according to the diagnostic line ratio scheme outlined by Fesen, Blair & Kirshner (1985). The red spectra from June 3, 2005 (position A) showed both strong [SII] lines at 6717 and 6731Å and strong [NII] relative to Hα together with a weak detection of [OI] at 6300Å. The ratio of [NII]/Hα=0.95 (Table 2) falls well inside the range expected for a SNR. Furthermore, the ratio of [SII]/Hα=0.76 is also well within the SNR range and strongly indicative of shock heated gas. The matching blue flux-calibrated spectrum exhibits weak Hβ and strong [OII] 3727Å but no obvious [OIII]. Using the Balmer decrement to establish extinction directly is problematic since the spectra were taken independently (though simultaneously) with the DBS. Figures 3 and 4 show the red and blue flux calibrated spectrum from the June 3, 2005 observations at slit position A.

The June 6, 2005 red spectra (position B) also show (Fig. 5) strong [NII] and [SII] lines relative to Hα with [NII]/Hα=1.90, [SII]/Hα=1.50 and [SII] 6717/6731=1.50. The corresponding blue spectra (Fig. 6) is broadly similar to the blue spectra from position A but this time [OIII] is detected at about twice Hβ. The ratios of all lines in the blue relative to Hβ=100 for both position A and B spectra are given in Table 3.

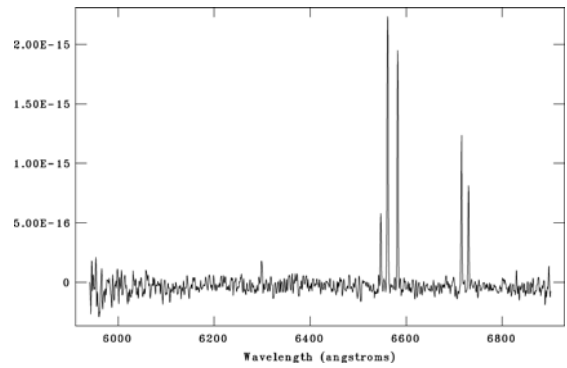

Fig. 3. 1-D, flux calibrated, high resolution, (1200 lines/mm grating) red spectrum of G315.1+2.7 for slit position A from June 3, 2005. All the usual prominent emission lines used to characterize an SNR are seen: [OI] at 6300Å, [NII] at 6548 and 6584Å, Hα and very strong [SII] at 6717 and 6731Å. One can note the relatively low intensity of [NII] 6548Å. The ratio of [NII]/Hα=1.1 still classifies G315.1+2.7 as a new Galactic SNR.

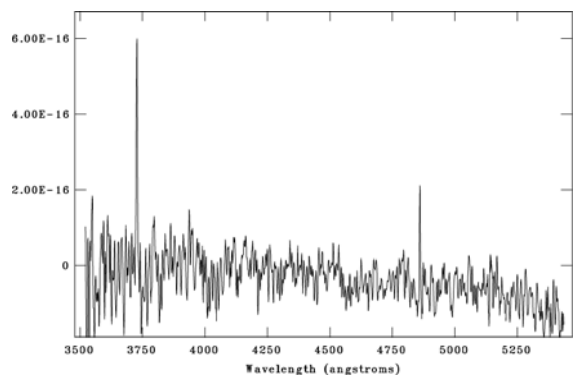

Fig. 4. 1-D, flux calibrated, medium resolution, (600 lines/mm grating) blue spectrum of G315.1+2.7 for slit position A from June 3, 2005. The [OIII] line appears absent though the Hβ line at 4861Å is evident and, very important for verification of new SNR, the [OII] 3727Å emission line.

---
[4] http://www.starlink.rl.ac.uk/

Table 2. Red fluxes relative to Hα for the two slit positions from Fig. 1 for G315.1+2.7. Measurements are from the high resolution, 1200 lines/mm grating.

| λ (Å) | Line | Flux (Hα = 100) | |
|---|---|---|---|
| | | Slit pos. A | Slit pos. B |
| 6300 | [OI] | 9 | 19 |
| 6364 | [OI] | 2 | -- |
| 6548 | [NII] | 21 | 50 |
| 6563 | Hα | 100 | 100 |
| 6584 | [NII] | 75 | 141 |
| 6717 | [SII] | 45 | 91 |
| 6731 | [SII] | 32 | 61 |
| Radial velocity[a] (kms$^{-1}$) | | -67±5 | -58±3 |
| Line Ratio | | | |
| [SII] 6717/6731 | | 1.41 | 1.50 |
| [NII]/Hα | | 0.95 | 1.90 |
| [SII]/Hα | | 0.76 | 1.50 |

[a] For the radial velocity estimates [NII], Hα and [SII] lines were used from the high dispersion red spectra.
Hα flux slit positions: A=6.62 × 10$^{-15}$; B=7.09 × 10$^{-15}$; in units of ergs cm$^{-2}$ s$^{-1}$ (uncorrected for interstellar reddening). Fluxes for the brightest lines are accurate to ±10%

Table 3. Blue fluxes relative to Hβ for the two slit positions for G315.1+2.7. The medium dispersion 600 lines/mm grating was used.

| λ (Å) | Line | Flux (Hβ = 100) | | | |
|---|---|---|---|---|---|
| | | Slit pos. A | | Slit pos. B | |
| | | F | I | F | I |
| 3727 | [OII] | 328 | 407 | 608 | 674 |
| 4101 | Hδ | 32 | 37 | 23 | 24 |
| 4340 | Hγ | 42 | 47 | 50 | 52 |
| 4861 | Hβ | 100 | 100 | 100 | 100 |
| 4959 | [OIII] | -- | -- | -- | |
| 5007 | [OIII] | -- | -- | 215 | 212 |
| Extinction (Hγ/Hβ): | | | | | |
| $c$ | | 0.37 | | 0.18 | |
| E(B-V) | | 0.25 | | 0.12 | |
| A$_v$ | | 0.79 | | 0.28 | |

Hβ flux slit positions: A=2.04 × 10$^{-15}$; B=1.11 × 10$^{-15}$; in units of ergs cm$^{-2}$ s$^{-1}$ (uncorrected for interstellar reddening). Fluxes for the brightest lines are accurate to ±10%

## 2.4 Difference between the optical spectra

In the red spectra from positions A and B the [OI] line at 6300Å is evident but weaker for position A. The matching [OI] 6364Å line, in lock-step with the [OI] 6300Å line at a ratio of 1/3 is hard to detect in either case being apparently absent at slit position A and very weak at slit position B. Unfortunately these lines are also very strong in the night-sky spectrum and are thus susceptible to sky-subtraction artefacts at the modest resolution used, even with best attempts at careful sky subtraction. Therefore none of these lines are used in any subsequent analysis.

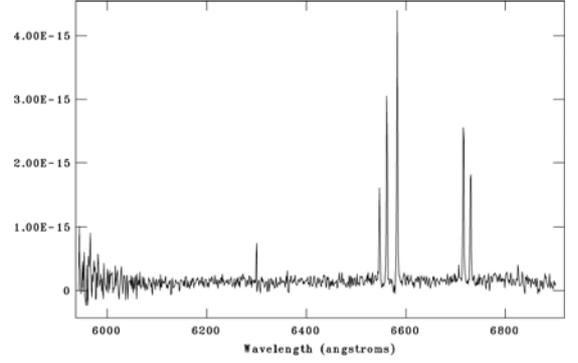

Fig. 5. 1-D, high resolution, flux calibrated, red spectrum of G315.1+2.7 for slit position B from June 6, 2005. Again, all prominent lines used to characterize an SNR are present: [OI] 6300Å, [NII] 6548 and 6584Å, Hα and very strong [SII] 6717 and 6731Å. A higher [NII]/Hα ratio can be noticed compared with the earlier spectrum from June 3, 2005.

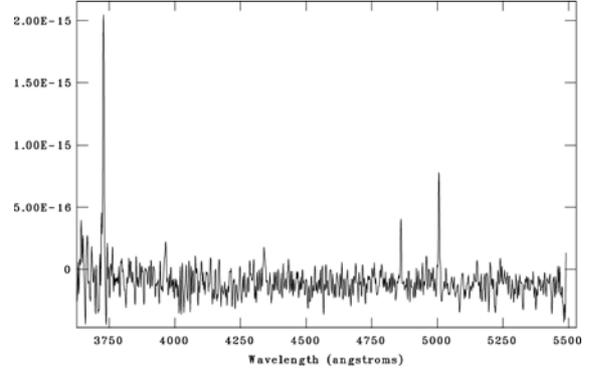

Fig. 6. 1-D, medium resolution, flux calibrated, blue spectrum (600 lines/mm grating) of G315.1+2.7 for position B from June 6, 2005. Again, the Hβ line at 4861Å can be recognized together with strong [OII] 3727Å. This spectrum is of generally weaker S/N than for the matching red spectrum but higher than for equivalent blue spectrum from position A. [OIII] at 5007Å is detected at about twice Hβ but the [OIII] 4959Å line is very weak.

Several other differences can be seen between the two sets of optical spectra, mainly in the varying strength of [NII] relative to Hα but also in the value of the [NII] 6584 to 6548Å ratio itself. Nevertheless, red spectra from slit position A and B both exhibit all the lines and ratios that strongly satisfy the criteria used for classification as a likely SNR and with a distinctly non HII region character: i.e. strong [NII] and [SII] relative to Hα with values >0.5 in both cases with the [SII]/Hα ratio being particularly telling (Fesen, Blair, & Kirshner 1985; Mathewson & Clarke 1973). Unusually however, the values of the [NII]/Hα and [SII]/Hα ratios from the two slit positions varies by ~100% which is much greater than normally exhibited between different locations across the same SNR as reported by Fesen, Blair, & Kirshner (1985). They typically find stability of these line ratios in individual remnants to be within 20% for [NII]/Hα and 30% for [SII]/Hα. Nevertheless it is quite clear that the new 10 arcmin filament is a distinct coherent structure. The [SII] 6717/6731 ratios (in fact emission-weighted average) from the two slit positions agree to within 5% well within the 20% range expected for within an individual SNR.

## 3 RADIO OBSERVATIONS

### 3.1 Parkes 2400 and PMN 4800 MHz observations

Duncan et al. (1995,1997) were the first to classify G315.1+2.7 as a Galactic SNR candidate. From their 2400 MHz Parkes radio observations they concluded that the candidate is a likely shell remnant seen "as an edge-brightened, broken ring or elliptical structure" 2.5° ✕ 3.2° in size. Their polarization maps also showed several bright patches of polarized emission, with an estimated overall radio flux at $S_{2.4GHz}$= 19±3 Jy. Fig.7, which is taken from the Parkes 2.4 GHz Sky Survey of the Southern Galactic Plane[5], shows the full extent of this object.

The medium resolution PMN 4850 MHz image, extracted from the archive (Fig. 8), shows G315.1+2.7 as a fractured elliptical shell which extends in the N-S direction and with the same dimension as for the associated, archived, lower resolution 2400 MHz images. The radio emission is missing only at the S-E arc of the shell remnant. Several point sources are also detected inside and on the rim of the remnant (with flux between 0.06 and 0.1 Jy beam$^{-1}$), but these are almost certainly unrelated background sources not connected with G315.1+2.7. A direct comparison with the new optical filament shows that the filament is exactly parallel to, and less than 1 arcmin from, the east edge of the radio source. This displacement is seen in close-up in Fig. 6.10 where contours for the matching higher resolution 843 MHz SUMSS radio data (Bock, Large & Sadler 1999) is overlaid on the Hα image.

G315.1+2.7 is an extremely large radio source, falling under the class of extended radio objects. The overall radio flux at 4850 MHz can be estimated but use of PMN fluxes has been shown to be problematic in many cases (Stupar et al. 2006, submitted). This is due to the effects of spillover temperature, the large number of overlaid background sources in the Galactic plane and the necessary subtraction of running-median baselines (e.g. Condon, Broderick & Seielstad 1989; Condon,

---

[5] http://www.uq.edu.au/~roy/data.html

Griffith & Wright 1993)

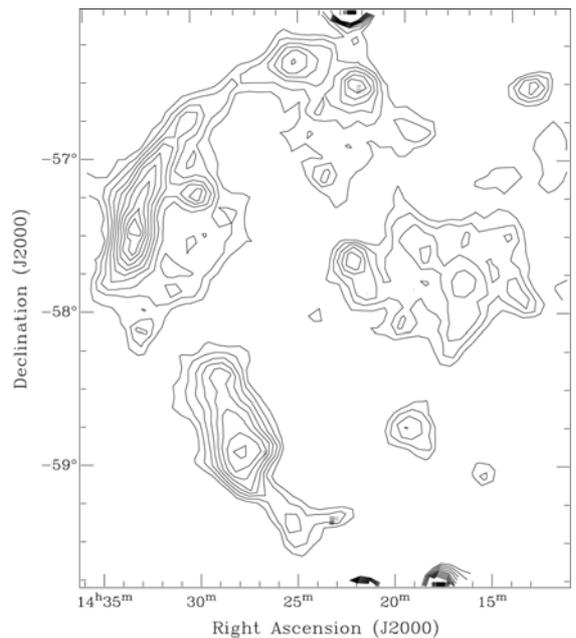

Fig. 7. G315.1+2.7 as seen at 2400 MHz and the basis for its preliminary identification as a Galactic SNR candidate by Duncat et al. (1997). The radio contours are between 0.14 and 0.5 Jy beam$^{-1}$ in steps of 0.04 Jy beam$^{-1}$. The image is from the Parkes 2400 MHz Sky Survey of the Southern Galactic Plane.

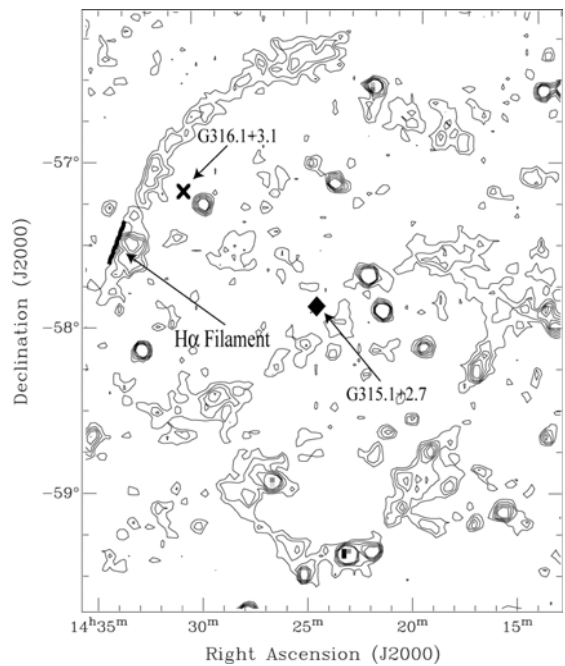

Fig. 8. G315.1+2.7 as seen at 4850 MHz from the PMN survey. The point-like sources evident inside the remnant are most likely background objects. Contours of the remnant's rim are between 0.01 and 0.05 Jy beam$^{-1}$ in steps of 0.01 Jy beam$^{-1}$. The Hα filament is shown at the end of the N-E arc. The ♦ sign shows the approximate centre of G315.1+2.7 and the ✕ sign marks the estimated centre for G316.1+3.12 (see later discussion).

## 3.2 SUMSS 843 MHz observations

The area of sky covering G315.1+2.7 is inside the publicly available SUMMS maps at 843 MHz. Part of the remnant can be identified at this frequency (Fig. 9), but not the whole extension, as seen clearly in the data at 4850 (PMN) and 2400 MHz. The only visible section of this remnant seen at 843 MHz is the N-E arc. However, due to the superior resolution of the SUMMS survey ($43" \times 43"$ csc$|\delta|$; see Bock, Large & Sadler 1999) compared to the PMN, improved structural details can be perceived (Figs. 9 and 10). There is an excellent positional match between the N-E arc seen in SUMSS and the equivalent PMN contours of the same arc at both frequencies.

As it is mentioned before, the position of the H$\alpha$ filament has also been directly compared with the arc seen in the SUMSS 843 MHz radio data where an even tighter positional coincidence is seen. It is clear that the optical filament closely follows the 843 MHz radio arc, partially overlapping on the central part (see Fig. 10).

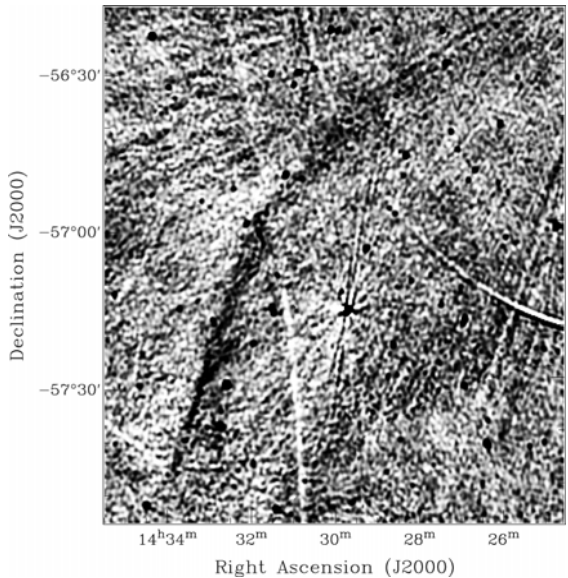

Fig. 9. Grayscale image of G315.1+2.7 as seen in the higher resolution 843 MHz SUMSS survey image. Comparing this image with the PMN 4850 MHz data shows that only the N-E arc of the remnant is evident in SUMSS and reproduced here. Several processing artefacts are also present in the 843 MHz data.

## 3.3 Other radio continuum observations

Combi, Romero & Arnal (1998) reported partial observations of G315.1+2.7 from within a large area of the sky surveyed at 1420 MHz in the constellation of Centaurus. They also took additional radio data from an all-sky survey at 408 MHz reported by Haslam et al. (1982). Their data were convolved and re-tabulated to the same beam as the 408 MHz maps. Using an estimated brightness temperature at both frequencies they produced a map of the spectral index distribution of this region. This map showed that G315.1+2.7 (e.g. G316.1+3.1) has a strong non-thermal spectral index of ~ -0.7 which is typical for an SNR.

In fact, Combi, Romero & Arnal (1998) gave Galactic coordinates for this SNR candidate that differed significantly from that reported by Duncan et al. (1995, 1997). They noticed 408 MHz peak emission at a different position, denoted by G316.1+3.1 (approximately R.A.= $14^h$ $30^m$ $47^s$ $\delta = -57°$ 12' 35", J2000). This is ~1 degree in the N-E direction from that given by Duncan et al. (1995, 1997) (see Fig. 8). This starkly illustrates that the peak frequency positions for the same large radio source can sometimes be different at different frequencies.

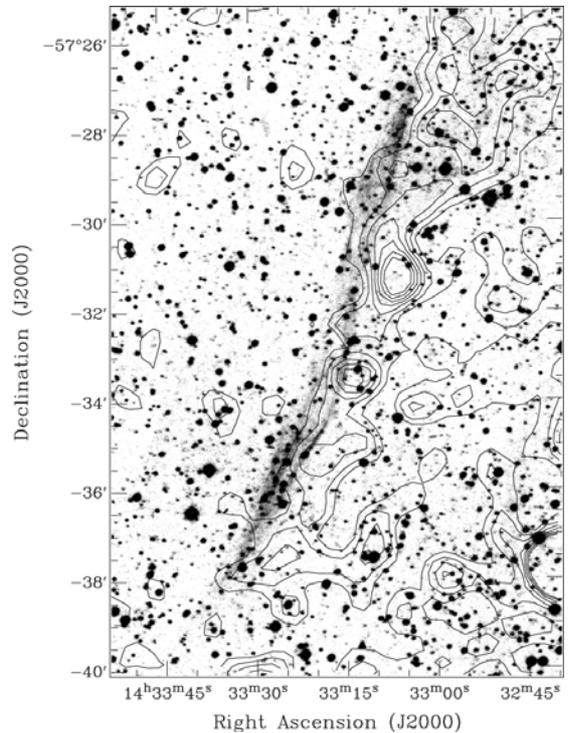

Fig. 10. G315.1+2.7 filament quotient image (H$\alpha$ and short red) overlaid and matched with SUMSS 843 MHz contours of the same field. Contours are from 0.001 to 0.01 in steps of 0.0015 Jy beam$^{-1}$. The displacement between the radio contours and optical filament is less than 1 arcmin.

## 4 SEARCH FOR POSSIBLE X-RAY SOURCE AND PULSAR

Detection of a central X-ray source and pulsar would lend additional strong credence to the identification of G315.1+2.7 as a bona-fide SNR. Unfortunately it is usually a very difficult task to search for reliable X-ray counterparts and especially for a pulsar (Camilo 2005) inside such a very large SNR as projected on the sky. Pulsars discovered inside or in the vicinity of an SNR may not always be connected with their parent remnant. They may have moved significantly from their central origin due to proper motion of both the star and remnant over the several tens of thousands of years since the original SN explosion. The same can be said for the connection of detected X-ray sources with Galactic SNRs. If the expansion of the remnant is uniform, and if the X-ray source connected with a given remnant exists, then it is usually detected around the central area of detected radio flux.

A SIMBAD[6] search around central area of G315.1+2.7 returned two ROSAT X-ray sources and three pulsars. Unfortunately, it is not possible to confirm the connection of any of these pulsars with this remnant. Two ROSAT sources, 1RXS J142112.2-575004 and 1RXS J142236.5-582748 are on the ROSAT list of bright sources (Voges et al. 1999). ROSAT source 1RXS J142112.2-575004 is closest to the approximate radio centre determined for G315.1+2.7 at a distance of some 30 arcmin. However, the other X-ray source 1RXS J142236.5-582748 is much brighter on the RASS broad band (0.1-2.4 keV; Voges et al. 1999) image with a diameter ~5.5 arcmin. Unfortunately, none of this information is sufficient to allow an unequivocal connection between the new remnant and either of these ROSAT bright X-ray sources to be made. Furthermore, checking explicitly for an X-ray source near the centre of G315.1+2.7 from the ROSAT All Sky Survey[7] and ROSAT Source Catalogs of Pointed Sources[8] did not return any palpable hits.

## 5 DISCUSSION

The intention of this paper is to classify the long-standing candidate SNR G315.1+2.7 as a bona-fide Galactic SNR. This conclusion is based on new, independent, optical imaging and subsequent, follow-up, spectroscopy that add credence to the identification based on existing multi-frequency radio data which we have collated. The position of the new optical filament discovered from the AAO/UKST Hα survey, correlates perfectly with the radio emission at and 843, 2400 and 4850 MHz. Furthermore, careful examination of the lower resolution SHASSA Hα data across the entire ~4°×4° region around the remnant's position, reveals a very low surface brightness, large-scale, ~3° arc structure which also seems to link directly to the wide-area radio maps at 2400 and 4850 MHz.

The observed [SII]/Hα is found to vary from 0.76 to 1.50 when sampled at two locations of the filament 9 arcmin apart but both these values well exceed the ratio found for HII regions, sitting squarely in the zone occupied by SNRs and confirming shock heated emission. The [NII]/Hα ratio also varies between values of 0.95 and 1.90 across the filament but likewise sits squarely outside any possible value for a HII region which is typically found to be <0.5 (see Fesen, Blair & Kirshner 1985). The filament is clearly not that of a PN based on the strength of the [SII] lines and presence of [OII] and [OIII] lines and in any event, unless very local, is far too large in angular extent. Few realistic alternatives to SNR identification seem plausible.

No significant observed variation of the sulphur doublet ratio ([SII] 6717/6731Å 1.41 and 1.50) was found inside the remnant, which is typically the case for an SNR (Fesen, Blair & Kirshner 1985). Additionally, the presence of several ionisation states of oxygen, strongly at 3727Å but weaker at 5007Å and 6300Å also support the SNR classification. Taken together, these new optical data and registration of G315.1+2.7 at 5 different radio frequencies, present a compelling picture that this source is a new, very large, galactic SNR, especially if the non-thermal nature of the radio flux is reliable. From the radio maps of brightness temperature at two frequencies (408 and 1420 MHz), Combi, Romero & Arnal (1998) confirmed the non-thermal nature of G315.1+2.7. However there is some ambiguity here as the spectral index (and accordingly radio flux at given frequencies) from Combi, Romero & Arnal (1998) disagrees with that inferred from the radio flux at 2400 MHz S=19±3 Jy estimated by Duncan et al. (1997).

Because the blue and the red spectra from the DBS on the 2.3m are obtained independently (although simultaneously), the usual ratio of flux calibrated Hα/Hβ to reliably estimate interstellar extinction needs to be done carefully (e.g. Oserbrock & Ferland 2006).

As an alternative and although the intensity of Hγ line is much lower, (and the S/N in the blue poorer) an extinction estimate was made based on the Hβ/Hγ ratio.

The relationship between observed and intrinsic intensities we used is

$$I_{obs}(H\gamma)/I_{obs}(H\beta) = I_{int}(H\gamma)/I_{int}(H\beta) 10^{-c[F(\lambda)]}$$

where $I_{obs}$ is observed ratio of Hγ/Hβ, $I_{int}$ is the intrinsic ratio Hγ/Hβ, $c$ is the extinction constant, i.e. the logarithmic extinction at Hβ and $F(\lambda)$ is the derived Galactic extinction function from Seaton (1979) which itself is based on Nandy et al (1975).

---

[6] http://simbad.u-strasbg.fr
[7] http://www.xray.mpe.mpg.de/cgi-bin/rosat/seq-browser
[8] http://www.xray.mpe.mpg.de/cgi-bin/rosat/rosat-survey

Our results for the estimated logarithmic correction $c$ were 0.37 and 0.18 for the two slit positions. This leads directly to E(B-V) estimates of 0.25 and 0.12 for E(B-V) (see Table 3). In addition, to the directly measured emission line fluxes, $F$, normalized to Hβ=100, we also derived extinction corrected line intensities ($I$) for interstellar extinction in the blue spectra (Table 3). As this correction is usually expressed as

$$I_{corr}(\lambda) = I_{obs}(\lambda) 10^{c[F(\lambda)]}$$

where $I_{corr}(\lambda)$ is the corrected (for interstellar extinction) and $I_{obs}(\lambda)$ is the observed line value normalized to Hβ we are unable to provide interstellar extinction correction for our red observations. Besides, taking into account that both observing nights were non-photometric and the fact that objects were observed not on high altitude but rather close to the horizon, this estimate of extinction should be taken with caution. Note that the factor of two difference in E(B-V) and $A_v$ may also reflect local variations in the density of the absorbing medium along the line of site to these two positions. Unfortunately examination of the IRAS 100 micron images do not reveal any significant differences in 100 micron flux that would arise from a higher dust continuum contribution to these two locations but this could simply reflect the poor resolution in the IRAS maps (~2 arcmin) compared with the size of G315.1+2.7 filament (~11 arcmin).

Due to insufficient resolution of the 600 lines/mm grating the [OII] doublet of 3726 and 3729Å in the blue spectrum is unresolved so we were unable to provide an estimate of electron density from this ratio (e.g. Osterbrock & Ferland 2006). Instead, the usual observed ratio of [SII] 6717/6731Å lines was used from the red spectra. Using an assumed electron temperature in this S$^+$ zone of non-equilibrium, shock recombination of ~10 000 K (Fesen, Blair & Kirshner 1985), we derived electron densities of ~33 cm$^{-3}$ for slit position A on Fig. 1.

A planar shock model of Hartigan, Raymond & Hartmann (1982) was used to estimate the shock velocity of ~100 kms$^{-1}$ for position B, only using the strength of [OIII] 5007Å line (this line is not seen in the spectrum at position A) relative to Hβ. Actually, to accurately determine shock velocity, we need UV lines (Hartigan, Raymond & Hartmann 1982), which are not covered, with our observations.

What is also interesting from the spectra at the B slit position is the very strong [NII] 6584Å emission compared with slit position A, where the [NII]/Hα ratio is typical for old or middle-age SNRs. We compared this strong nitrogen emission with the work of Blair, Long & Vancura (1991) where they confirmed [NII] 6584Å to Hα ratio of 1.6 on average for all knots they observed in Kepler's SNR. They concluded that a factor of 2 enhancement in N abundance was required to explain the observed line ratios where overabundance of nitrogen is due to the mass loss of the progenitor of Kepler's SNR. The fact that only at slit position B is extremely strong nitrogen observed, implies local enrichment is present and the Galactic N abundance gradient will not apply. However, for comparison, Stupar et al. (2006) spectroscopically observed new Galactic SNR G332.5-5.6 at five positions across its optical filaments seen in Hα light. Here, strong [NII] emission, as seen in Kepler's SNR, was found. The similarity between these observations and those of Kepler's SNR (Blair, Long & Vancura 1991), are clear. For the newly confirmed SNR G332.5-5.6, nitrogen enhancement is seen in all slit positions across the filaments with absolutely strong [NII] 6584Å and an average [NII]/Hα of ~2.42. This may indicate that the main optical filament of G315.1+2.7 presents an exceptional case in localised nitrogen variability.

Taking the relation:

$$n_{[SII]} \cong 45 n_c \times (V_c/100 kms^{-1})^2 (cm^{-3})$$

from Fesen & Krishner (1980) where $n_{[SII]}$ is the measured electron density of 33 cm$^{-3}$ derived from the position A [S II] ratio of 1.41 and $V_c$ is the shock velocity, the pre-shock cloud density was estimated to be 3 cm$^{-3}$. However, given that the position B [S II] ratio of 1.50 is higher than the low-density limit of 1.46 for $T_e = 10^4$ K, the overall values of $n_{[SII]}$ ranging from 1 to 66 cm$^{-3}$ are possible. In addition, we used the McKee & Cowie (1975) relation for blast wave energy, shock radius and shock cloud parameters:

$$E = 2 \times 10^{46} \beta^{-1} n_c (V_c/100 kms^{-1})^2 (r_s/1pc)^3 (erg)$$

to determine a radius $r_s$ of the remnant (in parsecs). In this equation, β is approximately equal to 1, and $E$ is the explosion energy in units of ergs.

Using the derived physical radius of the remnant from this equation and the observed mean radio angular diameter of 2.85 °, we can estimate a distance to G315.1+2.7 of 1.7 kpc on the basis of an assumed blast-wave energy E~10$^{51}$ ergs, though the uncertainties associated with the measured [S II] electron density allow distances from 1.4 to 5.4 kpc. Apart from the value of the mean electron density, the only other parameter that drastically changes this estimate is the blast explosion energy E which, according to theory, is between 10$^{50}$ and 10$^{51}$ ergs. If we take E=10$^{50}$ ergs, instead of 10$^{51}$, the distance to G315.1+2.7 reduces to ~0.9 kpc.

## 6 CONCLUSION

We have confirmed G315.1+2.7 as a bona-fide Galactic SNR based on a compelling set of new, optical Hα imagery and spectroscopy. This is positionally correlated with the extant multi-frequency radio data which we have re-evaluated and which exhibits a strong, negative, radio spectral index of ~ -0.7. The optical spectra, obtained at two well separated positions of the newly identified Hα filament, exhibit the pure spectral characteristics of an SNR. Both the radio and SHASSA Hα images have the form of a typical shell remnant with excellent positional coincidence between the radio and optical arcs (filaments). The distance has been determined on the basis of explosion energy ($E=10^{51}$) giving 1.7 kpc.

## ACKNOWLEDGMENTS

We are grateful to the referee Parviz Ghavamian for valuable comments and suggestions that have significantly improved the paper. One of the authors (MS) is thankful to the staff of the Wide Field Astronomy Unit at the Royal Observatory Edinburgh for their help during the visual inspection of original plates of the AAO/UKST Hα Survey of the Southern Galactic Plane in August 2004. We thank the MSSSO Time Allocation Committee for enabling the spectroscopic follow-up to be obtained. MS acknowledges the support of an APA PhD scholarship held at Macquarie University, Sydney, Australia.